\documentclass[onecolumn,secnumarabic,balancelastpage,amsmath,amssymb,nofootinbib]{article}

\usepackage[e]{esvect}
\usepackage{color}         
\usepackage{graphics}      
\usepackage{graphicx}      
\usepackage{epsf}          
\usepackage{bm}            
\usepackage[colorlinks=true]{hyperref}  

\usepackage{natbib}
\usepackage{amssymb}
\usepackage{amsmath}
\usepackage{mathrsfs}
\usepackage{framed}
\usepackage{bigints}

\definecolor{darkred}{rgb}{0.6,0,0}
\definecolor{darkgreen}{rgb}{0,0.5,0}
\definecolor{darkblue}{rgb}{0,0,0.6}
\hypersetup{ colorlinks,
linkcolor=darkblue,
filecolor=darkgreen,
urlcolor=darkgreen,
citecolor=darkred }

\begin{document}
\bibliographystyle{authordate1}


\title{Quantum Mechanics as Classical Physics}
\author{Charles T. Sebens\\University of Michigan, Department of Philosophy\\csebens@gmail.com\\ \small{[arXiv v.4]}}
\date{The published version appears in\\\textit{Philosophy of Science}, 82 (April 2015) pp. 266--291\\\small{\url{http://www.jstor.org/stable/10.1086/680190}}\\\small{submitted July 2013; revised August 2014}}

\maketitle

\begin{abstract}
Here I explore a novel no-collapse interpretation of quantum mechanics which combines aspects of two familiar and well-developed alternatives, Bohmian mechanics and the many-worlds interpretation.  Despite reproducing the empirical predictions of quantum mechanics, the theory looks surprisingly classical.  All there is at the fundamental level are particles interacting via Newtonian forces.  There is no wave function.  However, there are many worlds.
\end{abstract}


\section{Introduction}
On the face of it, quantum physics is nothing like classical physics.  Despite its oddity, work in the foundations of quantum theory has provided some palatable ways of understanding this strange quantum realm.  Most of our best theories take that story to include the existence of a very non-classical entity: the wave function.  Here I offer an alternative which combines elements of Bohmian mechanics and the many-worlds interpretation to form a theory in which there is no wave function.  According to this theory, all there is at the fundamental level are particles interacting via Newtonian forces.  In this sense, the theory is classical.  However, it is still undeniably strange as it posits the existence of a large but finite collection of worlds, each completely and utterly real.  When an experiment is conducted, every result with appreciable Born Rule probability does actually occur in one of these worlds.  Unlike the many worlds of the many-worlds interpretation, these worlds are fundamental, not emergent; they are interacting, not causally isolated; and they never branch.  In each of these worlds, particles follow well-defined trajectories and move as if they were being guided by a wave function in the familiar Bohmian way.

In this paper I will not attempt to argue that this theory is unequivocally superior to its competitors.  Instead, I would like to establish it as a surprisingly successful alternative which deserves attention and development, hopefully one day meriting inclusion among the list of promising realist responses to the measurement problem.

In \textsection \ref{measprob}, I briefly review why quantum mechanics is in need of a more precise formulation and discuss two no-collapse theories: the many-worlds interpretation and Bohmian mechanics.  I then go on to offer a rather unlikable variant of Bohmian mechanics which adds to the standard story a multitude of worlds all guided by the same wave function.  This theory is useful as a stepping stone on the way to Newtonian QM.  Newtonian QM is then introduced.  As soon as Newtonian QM is on the table, \textsection \ref{badstates} \& \ref{defining} present one of the most significant costs associated with the theory: the space of states must be restricted if the theory is to recover the experimental predictions of quantum mechanics.  In \textsection \ref{probEQM}, \ref{probBOHM}, \& \ref{probNQM}, I discuss the advantages of this new theory over Everettian and Bohmian quantum mechanics in explaining the connection between the squared amplitude of the wave function and probability.  In \textsection \ref{multitude}, I consider the possibility of modifying the theory so that it describes a continuous infinity of worlds instead of a finite collection, concluding that such a modification would be inadvisable.  In \textsection \ref{ontology}, I propose two options for the fundamental ontology of Newtonian QM.  In \textsection \ref{sym}, I use Newtonian QM to explain the way the wave function transforms under time reversal and Galilean boosts.  Spin is then discussed in \textsection \ref{withspin}.

Some limitations of the theory presented here are worth stating up front.  First, just as hydrodynamics relies on approximating a discrete collection of particles as a continuum, in its current form this theory must treat the discrete collection of worlds as a continuum.  As this is merely an approximation, empirical equivalence with standard quantum mechanics is likely only approximate (\textsection \ref{badstates}).  Second, one must impose a significant restriction on the space of states if the predictions of QM are to be reproduced (the Quantization Condition, \textsection \ref{defining}).  Third, I will not discuss extending the theory to handle multiple particles with spin or relativistic quantum physics.

\defcitealias{bostrom2012}{B\"{o}strom's}
\defcitealias{madelung1927}{Madelung's} 

Newtonian QM is a realist version of quantum mechanics based on the theory's hydrodynamic formulation (originally due to \citealp{madelung1927}).  For recent and relevant discussions of quantum hydrodynamics, see \citet{wyatt2005}; \citet{holland2005}.  An approach much like Newtonian QM was independently arrived at by \cite{HDW}.  Newtonian QM is somewhat similar to \citetalias{bostrom2012} \citeyearpar{bostrom2012} metaworld theory\footnote{The key difference with Newtonian QM being that B\"{o}strom's theory does not as thoroughly excise the wave function (the dynamics being given by \eqref{schrod} not \eqref{eom}).} and the proposal in \citet{tipler}.  Related ideas about how to remove the wave function are explored in \citet{poirier2010, schiff2012}, including a suggestion of many worlds.

To avoid confusion, throughout the paper I'll use ``universe'' to denote the entirety of reality, what philosophers call ``the actual world'' and what in these contexts is sometimes called the ``multiverse,'' reserving ``world'' for the many worlds of quantum mechanics.

\section{The Measurement Problem}\label{measprob}

If the state of the universe is given by a wave function and that wave function always evolves in accordance with the Schr\"{o}dinger equation, then quantum measurements will typically not have single definite outcomes.  Actual measurements of quantum systems performed in physics laboratories do seem to yield just one result.  This, in brief, is the measurement problem.  There are various ways of responding.

According to Everettian quantum mechanics, a.k.a.\ the many-worlds interpretation, the wave function $\Psi$ is all there is.  The evolution of the wave function is always given by the Schr\"{o}dinger equation,
\begin{equation}
i\hbar\frac{\partial}{\partial t}\Psi(\vv{x}_1,\vv{x}_2,...,t)
=\left(\sum_k{\frac{-\hbar^2}{2 m_k}\nabla_k^2}+V(\vv{x}_1,\vv{x}_2,...,t)\right)\Psi(\vv{x}_1,\vv{x}_2,...,t)\ ,
\label{schrod}
\end{equation}
where $\Psi$ is a function of particle configuration $(\vv{x}_1,\vv{x}_2,...)$ and time $t$, $m_k$ is the mass of particle $k$, $\nabla_k^2$ is the Laplacian with respect to $\vv{x}_k$, and $V$ is the classical potential energy of particle configuration $(\vv{x}_1,\vv{x}_2,...)$ at $t$.  When an observer performs a quantum measurement, the universal wave function enters a superposition of the observer seeing each possible outcome.  This is not to be understood as one observer seeing many outcomes, but as many observers each seeing a single outcome. Thus, the theory is not obviously inconsistent with our experience of measurements appearing to have unique outcomes.  According to Everettian quantum mechanics, there is nothing more than the wave function and therefore things like humans, measuring devices, and cats must be understood as being somehow composed of or arising out of wave function.  (\citealp{wallace2003, wallace2012} takes these things to be patterns or structures in the universal wave function.)  To summarize, here is what the Everettian QM says that there is (the ontology) and how it evolves in time (the dynamical laws).
\vspace*{6 pt}\\\hspace*{1.2cm}\textbf{Ontology:} (I) universal wave function $\Psi(\vv{x}_1,\vv{x}_2,...,t)$
\\\hspace*{1.2cm}\textbf{Law:} (I) Schr\"{o}dinger equation \eqref{schrod}
\vspace*{6 pt}

A second option in responding to the measurement problem is to expand the ontology so that the universe contains both a wave function evolving according to \eqref{schrod} \emph{and} particles with definite locations.  The time-dependent position of particle $k$ can be written as $\vv{x}_k(t)$ and its velocity as $\vv{v}_k(t)$.  The wave function pushes particles around by a specified law,
\begin{equation}
\vv{v}_k(t)=\frac{\hbar}{m_k}\mbox{Im}\left[\frac{\vv{\nabla}_k\Psi(\vv{x}_1,\vv{x}_2,...,t)}{\Psi(\vv{x}_1,\vv{x}_2,...,t)}\right]\ .
\label{guideq0}
\end{equation}
Experiments are guaranteed to have unique outcomes because humans and their scientific instruments are made of particles (not wave function).  These particles follow well-defined trajectories and are never in two places at once.  This theory is Bohmian mechanics, a.k.a.\ de Broglie-Bohm pilot wave theory.
\vspace*{6 pt}\\\hspace*{1.2cm}\textbf{Ontology:} (I) universal wave function $\Psi(\vv{x}_1,\vv{x}_2,...,t)$
\\\hspace*{1.2cm}(II) particles with positions $\vv{x}_k(t)$ and velocities $\vv{v}_k(t)$
\\\hspace*{1.2cm}\textbf{Laws:} (I) Schr\"{o}dinger equation \eqref{schrod}
\\\hspace*{1.2cm}(II) guidance equation \eqref{guideq0}\vspace*{6 pt}

From \eqref{schrod} and \eqref{guideq0}, one can derive an expression for the acceleration of each particle,
\begin{equation}
m_j \vv{a}_j(t)=-\vv{\nabla}_j\Big[Q(\vv{x}_1,\vv{x}_2,...,t)+V(\vv{x}_1,\vv{x}_2,...,t)\Big]\ ,
\label{oldeom}
\end{equation}
where $Q(\vv{x}_1,\vv{x}_2,...,t)$ is the quantum potential, defined by
\begin{equation}
Q(\vv{x}_1,\vv{x}_2,...,t)=\sum_k \frac{-\hbar^2}{2 m_k }\left(\frac{\nabla_k^2 |\Psi(\vv{x}_1,\vv{x}_2,...,t)|}{|\Psi(\vv{x}_1,\vv{x}_2,...,t)|}\right)\ .
\label{defofQ}
\end{equation}

Since the focus of this paper is not on Everettian or Bohmian quantum mechanics, I've sought to present each as simply as possible.  The best way to formulate each theory---ontology and laws---is a matter of current debate.

\section{Prodigal QM}

As a precursor to the theory I'll propose, consider the following interpretation of quantum mechanics which has both a many-worlds and a Bohmian flavor.  The wave function always obeys the Schr\"{o}dinger equation.  There are many different worlds, although a finite number, each represented by a point in configuration space\footnote{The location of a single particle is given by a point in space, $(\vv{x})$.  The locations of all particles are given by a point in \emph{configuration space}, $(\vv{x}_1,\vv{x}_2,...)$, where $\vv{x}_i$ is the location of particle $i$.}.  There are more worlds where $|\Psi|^2$ is large and less where it is small.  Each world is guided by the single universal wave function in accordance with the Bohmian guidance equation and thus each world follows a Bohmian trajectory through configuration space.  Let's call this ontologically extravagant theory \emph{Prodigal QM}.\footnote{With a continuous infinity of worlds, Prodigal QM is mentioned in \citet[][\textsection 7]{valentini2010} and in \citet{barrett1999} (in Barrett's terminology, it is a Bohmian many-threads theory in which all of the threads are taken to be completely real); a closely related proposal is discussed in \citet{dorr2009}.}  Why include a multitude of worlds when we only ever observe one, our own?  We could simplify the theory by removing all of the worlds but one, arriving at Bohmian mechanics \citep[][\textsection 7]{valentini2010}.  But, less obviously, it turns out that there is another route to simplification: keep the multitude of worlds but remove the wave function.  This option will be explored in the next section.

According to Prodigal QM, the universe contains a wave function $\Psi(\vv{x}_1,\vv{x}_2,...,t)$ on configuration space and a large number of worlds which can be represented as points moving around in configuration space.  The arrangement of the worlds in configuration space is described by a number density, $\rho(\vv{x}_1,\vv{x}_2,...,t)$, normalized so that integrating $\rho$ over all of configuration space gives one, $\int \! d^3 x_1 d^3 x_2...\: \rho=1$.  Integrating $\rho(\vv{x}_1,\vv{x}_2,...,t)$ over a not-too-small volume of configuration space gives the proportion of all of the worlds that happen to be in that volume at $t$.  By hypothesis, worlds are initially distributed so that
\begin{equation}
\rho(\vv{x}_1,\vv{x}_2,...,t)=|\Psi(\vv{x}_1,\vv{x}_2,...,t)|^2\ .
\label{probcon}
\end{equation}
The velocities of the particles are described by a collection of velocity fields indexed by particle number, $k$,
\begin{equation}
\vv{v}_k(\vv{x}_1,\vv{x}_2,...,t)=\frac{\hbar}{m_k}\mbox{Im}\left[\frac{\vv{\nabla}_k\Psi(\vv{x}_1,\vv{x}_2,...,t)}{\Psi(\vv{x}_1,\vv{x}_2,...,t)}\right]\ ,
\label{guideq}
\end{equation}
In Prodigal QM, if there is a world at $(\vv{x}_1,\vv{x}_2,...)$ at $t$ the velocity of the $k$th particle in that world is $\vv{v}_k(\vv{x}_1,\vv{x}_2,...,t)$.\footnote{This is not true for Newtonian QM (see \textsection \ref{badstates}).}  With these velocity fields, the equivariance property of the Bohmian guidance equation \eqref{guideq0} ensures that $\rho$ is always equal to $|\Psi|^2$ if it ever is \citep[see][\textsection 3]{durr1992}.
\vspace*{6 pt}\\\hspace*{1.2cm}\textbf{Ontology:} (I) universal wave function $\Psi(\vv{x}_1,\vv{x}_2,...,t)$
\\\hspace*{1.2cm}(II) particles in many worlds described by a world density $\rho(\vv{x}_1,\vv{x}_2,...,t)$ and velocity
\\\hspace*{1.2cm}fields $\vv{v}_k(\vv{x}_1,\vv{x}_2,...,t)$
\\\hspace*{1.2cm}\textbf{Laws:} (I) Schr\"{o}dinger equation \eqref{schrod}
\\\hspace*{1.2cm}(II) guidance equation \eqref{guideq}\footnote{Actually, the second dynamical law is more specific than \eqref{guideq} since it requires not just that the velocity \emph{fields} obey \eqref{guideq} but that each world follows an \emph{exact} Bohmian trajectory (see \textsection \ref{badstates}).  The connection between $\rho$ and $\Psi$ in \eqref{probcon}, though not a \emph{dynamical} law, might best be thought of as a third law of Prodigal QM.}\vspace*{6 pt}

The use of densities and velocity fields is familiar from fluid dynamics.  A quick review will be helpful.  Consider a fluid composed of $N$ point particles which each have mass $m$.  The number density of these particles is $n(\vv{x},t)$, normalized so that $\int \! d^3 x_1 d^3 x_2...\: n =N$.  The mass density is $m\!\times\! n(\vv{x},t)$.  Integrating $n(\vv{x},t)$ over a not-too-small volume gives the number of particles in that volume at $t$.  Whereas $n(\vv{x},t)$ gives the density of \emph{particles} in \emph{three}-dimensional space, $\rho$ gives the density of \emph{worlds} in \emph{configuration} space.  The velocity field for the fluid is $\vv{u}(\vv{x},t)$, defined as the mean velocity of particles near $\vv{x}$ at $t$.\footnote{More precisely, the number density and velocity field provide a good description of the particle trajectories if to a good approximation: $n(\vv{x},t)$ gives the average number of particles in a small-but-not-too-small region $\mathcal{R}$ centered about $\vv{x}$ over a short-but-not-too-short period of time $\mathcal{T}$ around $t$ divided by the volume of $\mathcal{R}$, and $\vv{u}(\vv{x},t)$ gives the average velocities of the particles in $\mathcal{R}$ over $\mathcal{T}$.  For more detail, see \citet[][\textsection 2.2]{chapman1970}.  The connection between $\rho$ and the $\vv{v}_k$s and the trajectories of individual worlds could be spelled out along similar lines, but full rigor in the context of Newtonian QM would require a better understanding of the dynamics (see \textsection \ref{badstates} and \citealp{HDW}).\label{morecareful}}  For an inviscid compressible fluid with zero vorticity, the time evolution of $n$ and $\vv{u}$ are determined by a continuity equation
\begin{equation}
\frac{\partial n(\vv{x},t)}{\partial t}=-\vv{\nabla} \cdot \Big(n\left(\vv{x},t\right)\vv{u}\left(\vv{x},t\right)\Big)\ ,
\label{fluidhamster}
\end{equation}
and a Newtonian force law
\begin{equation}
m \vv{a}(\vv{x},t)=-\vv{\nabla}\left[\frac{p(\vv{x},t)}{n(\vv{x},t)}+V\left(\vv{x},t\right)\right]\ ,
\label{fluideom}
\end{equation}
where $V$ is the external potential, $p$ is the pressure, and
\begin{equation}
\vv{a}(\vv{x},t)=\frac{D\vv{u}(\vv{x},t)}{Dt}=\left(\vv{u}(\vv{x},t)\cdot\vv{\nabla}\right)\vv{u}(\vv{x},t)+\frac{\partial \vv{u}(\vv{x},t)}{\partial t}\ .
\label{fluidaccel}
\end{equation}
The acceleration is given by the material derivative of $\vv{u}$ not the partial derivative because a particle's position in the fluid is time dependent.

\begin{figure}[h!]
\centerline{\includegraphics[width=6cm]{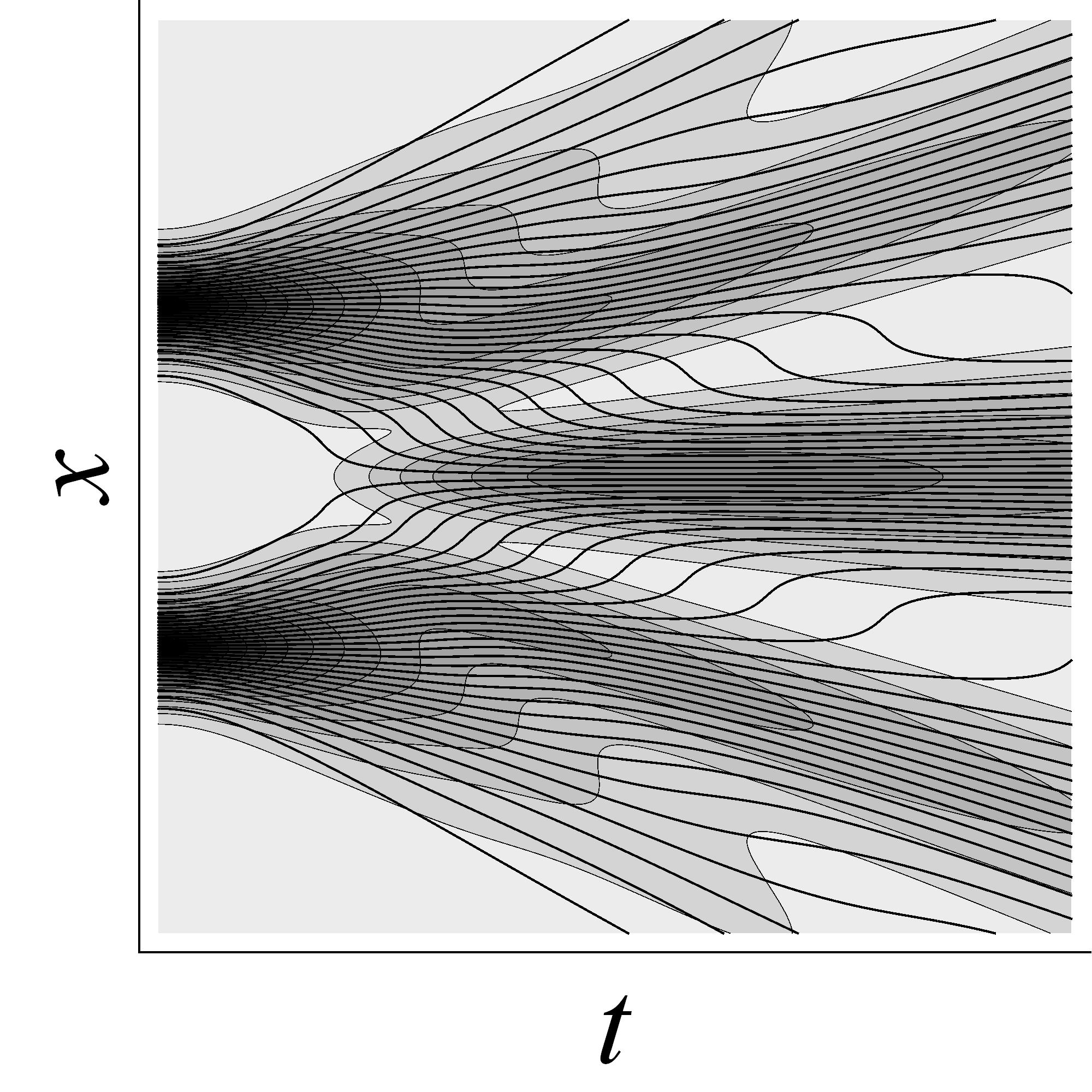}\ \ \ \ \ \ \includegraphics[width=6cm]{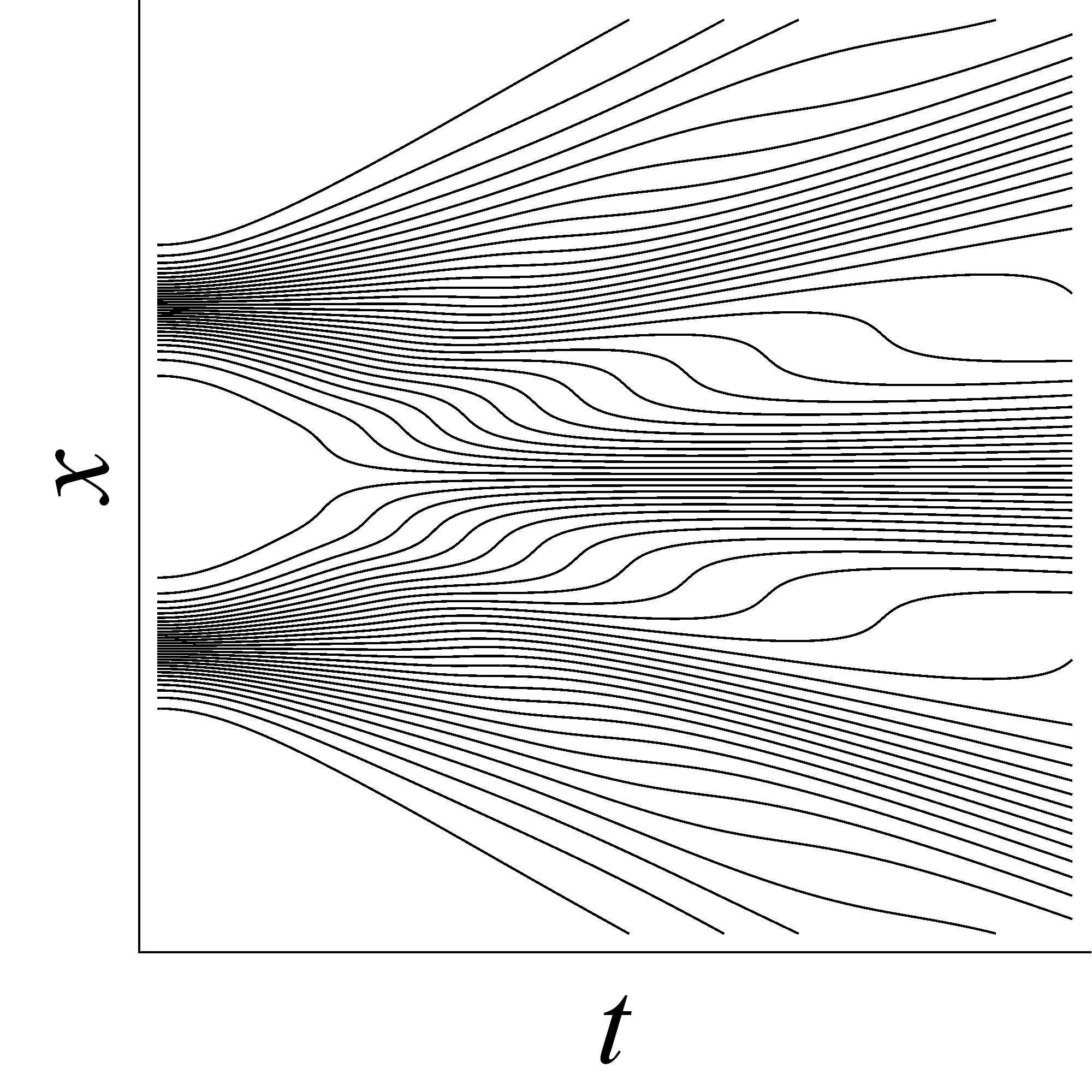}}
\vspace*{-4 pt}
\large{\hspace*{3 cm} Prodigal QM \hspace{3.9 cm} Newtonian QM}\vspace*{8 pt}\\
\centerline{\includegraphics[width=6cm]{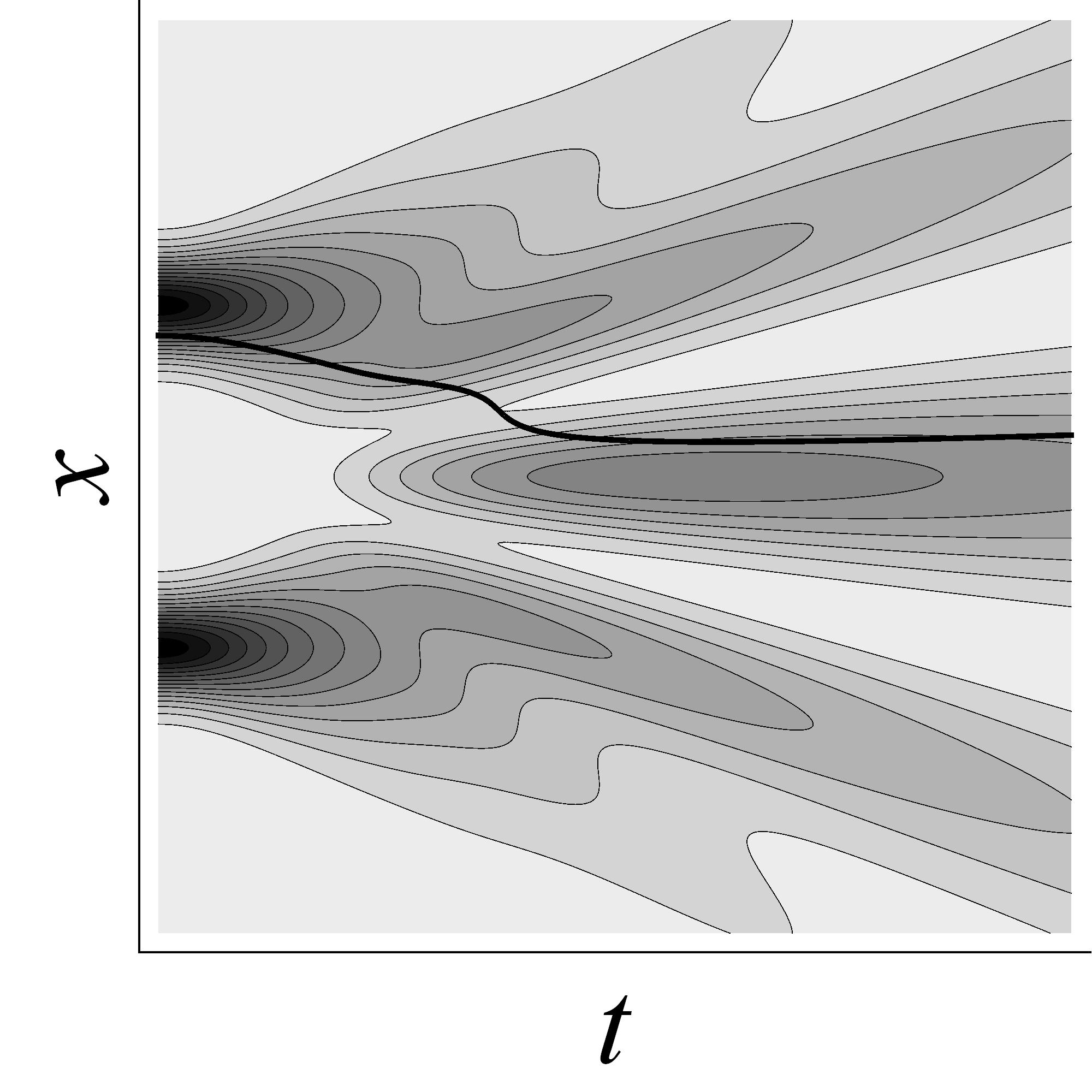}\ \ \ \ \ \ \includegraphics[width=6cm]{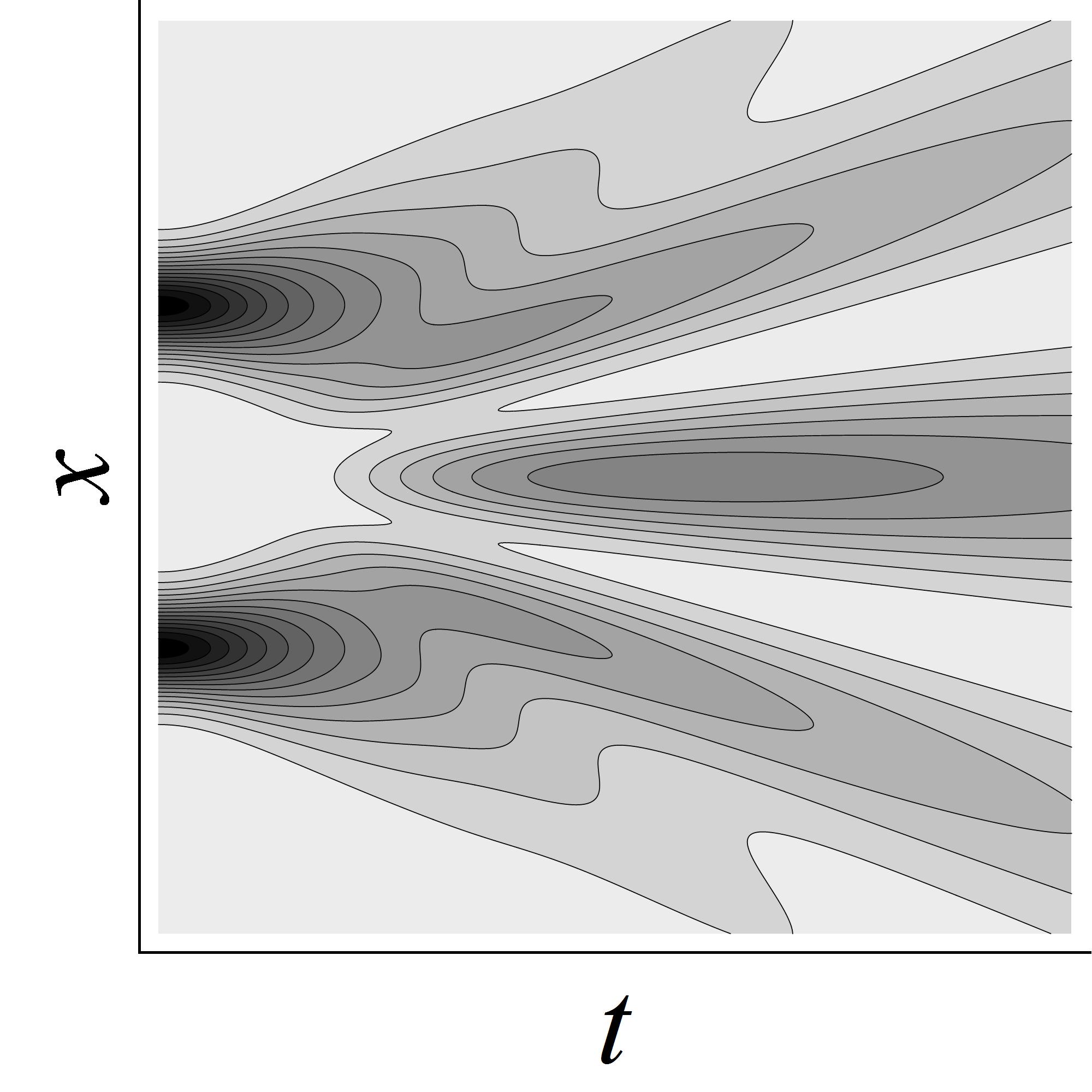}}\vspace*{-4 pt}
\large{\hspace*{3 cm} Bohmian QM \hspace{3.9 cm} Everettian QM}
\caption{Diagrams of the evolution of a single particle in the double-slit experiment according to four different no-collapse theories.  The vertical axis gives the position of the single particle and the horizontal axis time.  $|\Psi|^2$ is shown as a contour plot and particle trajectories as lines.}
\label{NoCollapseQMs}
\end{figure}

The three quantum theories on the table thus far are applied to the double-slit experiment in figure \ref{NoCollapseQMs}.  In the bottom-right diagram is Everettian QM where the universe is just a wave function.  The particle's wave function is initially peaked at the two slits and then spreads out and interferes as time progresses.  When the particle hits the detector, a multitude of worlds will separate via decoherence and in each the particle will be observed hitting at a particular point on the screen.  In Bohmian mechanics, one adds to the wave function an actual particle which follows a definite trajectory in accordance with the guidance equation.  In Prodigal QM, there is a wave function \emph{and} a collection of worlds, each of which contains a particle following a Bohmian trajectory.  In Newtonian QM, which will be introduced at the end of \textsection \ref{removing}, one retains the multitude of worlds but removes the wave function.

\section{Removing the Wave Function}\label{removing}

One can derive an equation for the dynamics of particles in Prodigal QM that makes no reference to the wave function.  Once this is done, we can formulate an alternate theory where the superfluous wave function has been removed.  This new theory, Newtonian QM, will be the focus of the remainder of the article.  The mathematical manipulations presented in this section are familiar from discussions of Bohmian mechanics, but take on a different meaning as derivations of particle dynamics in Prodigal QM.  Those who wish to skip the derivation should simply note that \eqref{eom} is derivable from \eqref{schrod}, \eqref{probcon}, and \eqref{guideq}.

As $\rho=|\Psi|^2$ \eqref{probcon}, the wave function can be written in terms of the world-density and a phase factor as
\begin{equation}
\Psi(\vv{x}_1,\vv{x}_2,...,t)=\sqrt{\rho(\vv{x}_1,\vv{x}_2,...,t)}e^{i\theta(\vv{x}_1,\vv{x}_2,...,t)}\ .
\label{first}
\end{equation}
Plugging \eqref{first} into the guidance equation \eqref{guideq} generates
\begin{equation}
\vv{v}_k(\vv{x}_1,\vv{x}_2,...,t)=\frac{\hbar}{m_k}\vv{\nabla}_k\theta(\vv{x}_1,\vv{x}_2,...,t)\ ,
\label{guideq2}
\end{equation}
relating $\vv{v}_k$ and $\theta$.  (At this point, I will stop repeating the arguments of $\Psi$, $\rho$, $\theta$, and $\vv{v}_k$; they \emph{all} depend on the configuration of particles and the time.)

The evolution of the wave function $\Psi$ is given by the Schr\"{o}dinger equation \eqref{schrod}.  Dividing both sides of \eqref{schrod} by $\Psi$ and using \eqref{first}, one can derive that
\begin{equation}
\frac{i\hbar}{2\rho}\frac{\partial\rho}{\partial t}-\hbar \frac{\partial \theta}{\partial t} = \sum_k{\frac{-\hbar^2}{2 m_k}\left[\frac{\nabla_k^2 \sqrt{\rho}}{\sqrt{\rho}}+\frac{2 i}{\sqrt{\rho}} \left(\vv{\nabla}_k \sqrt{\rho}\right)\cdot\left(\vv{\nabla}_k \theta\right) + i \: \nabla_k^2 \theta - \left|\vv{\nabla}_k \theta\right|^2\right]}+V\ .
\label{astep}
\end{equation}
Equating the imaginary parts, using \eqref{guideq2}, yields
\begin{equation}
\frac{\partial\rho}{\partial t}=-\sum_k{\vv{\nabla}_k \cdot \left(\rho\vv{v}_k\right)}\ ,
\label{hamster}
\end{equation}
a continuity equation similar to \eqref{fluidhamster}.  Equating the \emph{real} parts of \eqref{astep}, using \eqref{guideq2}, yields
\begin{equation}
\frac{\partial \theta}{\partial t}=\sum_k{\left\{\frac{\hbar}{2 m_k}\frac{\nabla_k^2 \sqrt{\rho}}{\sqrt{\rho}}-\frac{m_k}{2 \hbar}|\vv{v}_k|^2\right\}}-\frac{V}{\hbar}\ .
\label{thetaT}
\end{equation}
Acting with $\frac{\hbar}{m_j}\vv{\nabla}_j$ on both sides of \eqref{thetaT} and rearranging, making use of \eqref{guideq2} and the fact that
\begin{equation}
\vv{a}_j=\sum_k (\vv{v}_k\cdot\vv{\nabla}_k)\vv{v}_j+\frac{\partial \vv{v}_j}{\partial t}
\label{accel}
\end{equation}
gives
\begin{equation}
m_j \vv{a}_j=-\vv{\nabla}_j\left[\sum_k \frac{-\hbar^2}{2 m_k }\left(\frac{\nabla_k^2 \sqrt{\rho}}{\sqrt{\rho}}\right)+V\right]\ .
\label{eom}
\end{equation}
We have derived an equation of motion of the form $F=ma$, similar to both \eqref{oldeom} and \eqref{fluideom}.\footnote{This is the multi-particle version of \citet[][eq. 1.7]{wyatt2005}; \citet[][eq. 4.9]{holland2005}.}  The last term in the brackets gives the \emph{classical} potential energy of the configuration of particles and makes no reference to the other worlds.  The other term looks like an \emph{interaction} between the worlds. This term is the quantum potential $Q$ familiar from Bohmian mechanics \eqref{defofQ}, with $|\Psi|$ replaced by $\sqrt{\rho}$.

Within Prodigal QM, we've seen that one can derive an equation which determines the dynamics for all of the particles in all of the worlds \emph{without ever referencing the wave function}.  \eqref{eom} gives a way of calculating the acceleration of a particle that doesn't mention $\Psi$, as \eqref{guideq} does, but only depends on the density of worlds $\rho$ and the potential $V$.  In Prodigal QM, this equation is derived, not part of the statement of the theory in the previous section.  But, what if we took it to be the primary equation of motion for the particles?  One can remove the wave function from Prodigal QM leaving only the corresponding $\rho$ and $\vv{v}_k$s.  So long as one enforces \eqref{eom}, the dynamics for particles will be essentially as they were in Prodigal QM.

Now we can formulate a new theory: \emph{Newtonian QM}.  Reality consists of a large but finite number of worlds whose distribution in configuration space is described by $\rho(\vv{x}_1,\vv{x}_2,...,t)$.  The velocities of the particles in the worlds are described by the velocity fields $\vv{v}_k(\vv{x}_1,\vv{x}_2,...,t)$.  The dynamical law for the velocity fields is \eqref{eom}, a Newtonian force law.  As the particles move, the resultant shift in the distribution $\rho$ is determined by \eqref{hamster}.  According to Newtonian QM, quantum mechanics is nothing but the Newtonian mechanics of particles in many different worlds.
\vspace*{6 pt}\\\hspace*{1.2cm}\textbf{Ontology:} (I) particles in many worlds described by a world density $\rho(\vv{x}_1,\vv{x}_2,...,t)$
\\\hspace*{1.2cm}and velocity fields $\vv{v}_k(\vv{x}_1,\vv{x}_2,...,t)$
\\\hspace*{1.2cm}\textbf{Law:} (I) Newtonian force law \eqref{eom}\footnote{The continuity equation \eqref{hamster}, although used alongside \eqref{eom} to calculate the dynamics, is not here considered a dynamical \emph{law} since it merely encodes the fact that worlds are neither created nor destroyed.  As is mentioned in \textsection \ref{defining}, the Quantization Condition might be considered a non-dynamical law.}\vspace*{6 pt}

Comparing this statement of Newtonian QM to the formulation of Bohmian mechanics in \textsection \ref{measprob}, Newtonian QM is arguably the simpler theory.  The theory has a single dynamical law and the fundamental ontology consists only of particles.  However, this quick verdict could certainly be contested, especially in light of the discussion below: \eqref{eom} is not a fundamental law (\textsection \ref{badstates}); an unnatural restriction must be put on the space of states (\textsection \ref{defining}); there are multiple ways to precisify ontology of the theory (\textsection \ref{ontology}).

\section{The Continuum Approximation}\label{badstates}

Since the number of worlds is taken to be finite, the actual distribution of worlds will be highly discontinuous; some locations in configuration space will contain worlds and others will not.  Still, we can use a smooth density function $\rho$ to describe the distribution of worlds well enough at a coarse-grained level (see footnote \ref{morecareful}).  The velocity field $\vv{v}_k(\vv{x}_1,\vv{x}_2,...)$ gives the mean velocity of the $k$-th particle in worlds near $(\vv{x}_1,\vv{x}_2,...)$, but the $k$-th particle in a world at $(\vv{x}_1,\vv{x}_2,...)$ may have a somewhat different velocity from $\vv{v}_k(\vv{x}_1,\vv{x}_2,...)$.  So, in Newtonian QM worlds will typically only approximately follow Bohmian trajectories through configuration space just as fluid particles do not exactly follow pathlines.\footnote{A \emph{pathline} gives the trajectory of a particle always traveling at the mean velocity $\vv{u}$.}

In fluid dynamics, the use of a description of the fluid in terms of $n$ and $\vv{u}$ is justified by the fact that we can calculate the dynamics of these coarse-grained properties (and others) without needing to know exactly what all the particles are doing.  Also, it is the coarse-grained properties that we measure (\citealp[\textsection 1.2]{batchelor1967}; \citealp[\textsection 5]{chapman1970}).  What justifies the use of $\rho$ and the $\vv{v}_k$s to describe the collection of worlds?  As it turns out, we can calculate the dynamics of these properties without worrying about the exact locations of worlds via \eqref{hamster} and \eqref{eom}.  Once the evolution of $\rho$ and the $\vv{v}_k$s are known, we can use $\rho(t)$ to get probabilities (\textsection \ref{probNQM}) and the $\vv{v}_k(t)$s to determine pathlines (showing that particles follow Bohmian trajectories).

The equation of motion for the theory \eqref{eom} treats the collection of worlds as a continuum.  It fails to be a fundamental law since it does not describe the precise evolution of each world and is not valid if there are too few worlds to be well-described as a continuum.  Slight deviations from standard quantum mechanical behavior should be expected due to the fact that there are only a finite number of worlds; worse deviations the fewer worlds there are.  Future experiments may observe such deviations and support Newtonian QM.  As textbook quantum mechanics works well, we have reason to believe there are a very large number of worlds.  (The situation here is similar to that of spontaneous collapse theories, which are in principle empirically testable.)  Ultimately, the quantum contribution to the force in \eqref{eom} should be derivable from a more fundamental inter-world interaction.  One should be able to calculate the forces when there are only a handful of worlds.  Hopefully future research will explain how the continuum approximation arises from a ``micro-dynamics'' of worlds just as fluid dynamics arises from the micro-dynamics of molecules.  For some progress in this direction, see \citet{HDW}.

\section{Reintroducing the Wave Function}\label{defining}

In \textsection \ref{removing} we saw that for any wave function $\Psi(t)$ obeying the Schr\"{o}dinger equation, there exists a world-density $\rho(t)$ and a collection of velocity fields $\vv{v}_k(t)$ obeying \eqref{eom} such that the relations between $\Psi$, $\rho$, and the $\vv{v}_k$s expressed in \eqref{probcon} and \eqref{guideq} are satisfied at all times.  The converse does not hold.  There are some combinations of $\rho$ and the $\vv{v}_k$s, that is, some ways the universe might be according to Newtonian QM, that do not correspond to any wave function.  In general, we'll restrict our attention to combinations of $\rho$ and the $\vv{v}_k$s that can be derived from a wave function via \eqref{probcon} and \eqref{guideq} as it is these states which reproduce the predictions of quantum physics.  For such states, it may be useful to \emph{introduce} a wave function, $\Psi$, even though it is not a fundamental entity and does not appear in the equation of motion of the theory \eqref{eom}.  The wave function serves as a convenient way of \emph{summarizing} information about the positions and velocities of particles in the various worlds; the magnitude encodes the density of worlds \eqref{probcon} and the phase encodes the velocities of particles \eqref{guideq2}.  The wave function need not be mentioned in stating the theory or (in principle) for deriving empirical predictions, but introducing a wave function is useful for making contact with standard treatments of quantum mechanics.

As was just mentioned, there are some states of the universe in Newtonian QM that do not correspond to quantum wave functions.\footnote{This point was made concisely by \citet{wallstrom1994} in the context of quantum hydrodynamics; it was noted earlier by \citet{takabayasi1952}; see also \citet[][eq. 4.14]{holland2005}.}  That is, there are some combinations of $\rho$ and the $\vv{v}_k$s for which one cannot find a wave function $\Psi$ that satisfies \eqref{probcon} and \eqref{guideq}.  The amplitude of $\Psi$ follows straightforwardly from $\rho$, but not every set of velocity fields $\vv{v}_k$ can be expressed as $\frac{\hbar}{m_k}$ times the gradient of a phase \eqref{guideq2}.  For this to be the case, we must impose a constraint on the velocity fields.\footnote{This is loosely analogous to the constraint on the fluid velocity field $\vv{u}$ that it be irrotational (everywhere zero vorticity) which is required to introduce a velocity potential (and for the validity of \eqref{fluideom}).}
\begin{description}
\item[Quantization Condition]  Integrating the momenta of the particles along any closed loop in configuration space gives a multiple of Planck's constant, $h=2\pi\hbar$.
\begin{equation}
\oint{\left\{\sum_k{\left[m_k\vv{v}_k \cdot d\vv{\ell}_k\right]}\right\}}=nh\ .
\label{requirement}
\end{equation}
\end{description}
If the Quantization Condition is satisfied initially, \eqref{eom} ensures that it will be satisfied at all times.

To see one sort of constraint this requirement imposes, think about the following case: a single electron orbiting a hydrogen nucleus in the $n=2$, $l=1$, $m=1$ energy eigenstate.  For simplicity, take the nucleus to provide an external potential and the universe to contain many worlds with a single electron in each.  The electron's wave function is
\begin{equation}
\Psi_{2,1,1}(r,\theta,\phi)=\frac{-1}{8 \sqrt{a^5 \pi}}e^{\frac{-r}{2a}}e^{i\phi}r\sin \theta\ ,
\label{circlingelectrons}
\end{equation}
where $a$ is the Bohr radius.  The guidance equation tells us that the particle in each world executes a circle around the $z$-axis with velocity $v_{\phi}=\frac{\hbar}{m r sin{\theta}}$, entirely in the $\widehat{\phi}$ direction (here $\phi$ is the azimuthal angle).  \eqref{requirement} is trivially satisfied since $\frac{m \hbar}{m r sin{\theta}} \times 2 \pi r sin{\theta}=h$.  But, if the electrons were circling the $z$-axis a bit faster or a bit slower the integral wouldn't turn out right and \eqref{requirement} wouldn't be satisfied; they could orbit twice as fast but not $1.5$ times as fast.

Without the Quantization Condition, Newtonian QM has too large a space of states.  There are ways the universe might be that are quantum mechanical and others that are not.  It is easy to specify what universes should be excluded, those that violate \eqref{requirement}, but hard to give a principled reason why those states should be counted as un-physical, improbable, or otherwise ignorable.  For now, I think it is best to understand the Quantization Condition as an empirically discovered feature of the current state of the universe, or equivalently, of the initial conditions.  However, one might prefer to think of it as a non-dynamical law.  A better explanation of the Quantization Condition's satisfaction would help strengthen Newtonian QM as it might seem that the best possible explanation of the condition's satisfaction is the \emph{existence} of a wave function (backtracking to Prodigal QM).  In the remainder of the article I will assume that the Quantization Condition is satisfied.

Suppose the world density and the velocity fields at a time are given.  Provided the Quantization Condition is satisfied, there exists a wave function satisfying \eqref{probcon} and \eqref{guideq}.  But, is it unique?\footnote{Here the question is considered at the level of the continuum description.  Because there are multiple ways of coarse-graining, there will be multiple not-too-different $\rho$s and $\vv{v}_k$s that well-describe any \emph{finite} collection of worlds and thus many wave functions.  It may be that some ways of coarse-graining avoid the problems raised below by ensuring that the velocity fields are always well-defined.  If they do, the derivability of $\Psi$ from $\rho$ and the $\vv{v}_k$s comes at the cost of limiting the wave functions one can recover, losing those in \eqref{circlingelectrons}, \eqref{trouble}, and \eqref{evil}.}  That is, can \eqref{probcon} and \eqref{guideq} be used to define $\Psi$ in terms of $\rho$ and the $\vv{v}_k$s?\footnote{See also the discussion in \citet[][\textsection 4]{holland2005}.}  First consider the case where $\rho$ is everywhere nonzero.  The magnitude of $\Psi$ can be derived from \eqref{probcon}, and \eqref{guideq2} gives the phase up to a global constant.  The wave function can be determined up to a global phase.  This would be insufficient if the overall phase mattered, but as the global phase is arbitrary this gives exactly what we need.  Actually, it's even better this way.  The fact that the dynamics don't care about the overall phase is explained in Newtonian QM by the fact that changes in the global phase of the wave function don't change the state of the universe; that is, they don't change $\rho$ or the $\vv{v}_k$s.

If the region in which $\rho\neq0$ is not connected, the wave function is not uniquely determined by $\rho$ and the $\vv{v}_k$s---one can introduce arbitrary phase differences between the separate regions.  As an example of the breakdown of uniqueness, consider the second energy eigenstate of a single particle in a one-dimensional infinite square well of length $L$.  In this case the wave function is
\begin{equation}
\psi_a(x)=\sqrt{\frac{2}{L}}\sin\left(\frac{2\pi x}{L}\right)\ .
\label{trouble}
\end{equation}
This describes a universe with $\rho$ and $\vv{v}$ given by
\begin{align}
\rho(x)&=\frac{2}{L}\sin^2\left(\frac{2\pi x}{L}\right)
\nonumber
\\
\vv{v}(x)&=\left\{\begin{array}{ll} 0 & \mbox{   if }x\neq \frac{L}{2} \\ undefined & \mbox{   if }x= \frac{L}{2} \end{array}\right.\ .
\label{dwbwoe}
\end{align}
The velocity field $\vv{v}$ is undefined where there are no worlds.  These expressions for $\rho$ and $\vv{v}$ are also compatible with\footnote{The wave function $\psi_b$ has the disreputable property of not being smooth.  It should be noted that there exist pairs of distinct smooth non-analytic wave functions which agree on $\rho$ and $\vv{v}$ at a time. (Thanks to Gordon Belot for suggesting an example like this.)  For example, 
\begin{align}
\psi_{\alpha}(x)=&\left\{\begin{array}{ll}
C e^{\frac{-1}{1-(x+2)^2}} & \mbox{       if }-3<x<-1 \\ -C e^{\frac{-1}{1-(x-2)^2}} & \mbox{       if }1<x<3 \\ 0 & \mbox{       else}\end{array}\right.
\nonumber
\\
\psi_{\beta}(x)=& \  |\psi_{\alpha}(x)| \ .
\nonumber
\end{align}}

\begin{equation}
\psi_b(x)=\sqrt{\frac{2}{L}}\left|\sin\left(\frac{2\pi x}{L}\right)\right| \ .
\label{evil}
\end{equation}

This exposes an inconvenient indeterminism: The time evolution of $\psi_a$ is trivial as it is an energy eigenstate.  Since $\psi_b$ is not differentiable at $L/2$, its time evolution cannot be calculated straightforwardly using the Schr\"{o}dinger equation \eqref{schrod}.  As  \eqref{probcon} and \eqref{guideq} do not determine which wave function is to be used to describe the state in \eqref{dwbwoe}, it is not clear how the state will evolve.  The future evolution of the universe is not uniquely determined by the instantaneous state \eqref{dwbwoe}, the continuity equation \eqref{hamster}, and the equation of motion \eqref{eom}.  This indeterminacy arises because $\rho$ is zero and the velocity field is undefined at $L/2$, so $\frac{\partial \rho}{\partial t}$ and $\vv{a}$ are undefined at $L/2$.  There is reason to think this indeterminism is an artifact of the continuum approximation where \eqref{hamster} and \eqref{eom} need the velocity fields to be well-defined at every point in configuration space---even where there are no worlds---to yield a unique time evolution. The fundamental dynamics should take as input a specification of the position of each world in configuration space and the velocities of the particles in those worlds, all of which will be well-defined (\textsection \ref{badstates}).

Consider a slightly different problem from that just considered: Suppose one would would like to find a wave function $\Psi(t)$ which describes a \emph{history} of $\rho(t)$ and the $\vv{v}_k(t)$s, satisfying \eqref{hamster} and \eqref{eom} over some time interval.  There will be a collection of wave functions which satisfy \eqref{probcon} and \eqref{guideq} at each time.  For any such wave function, one can multiply it by a spatially homogeneous time-dependent phase factor, $e^{if(t)}$, to get another wave function which always satisfies \eqref{probcon} and \eqref{guideq}.  (The global phase at each time is arbitrary and \eqref{probcon} and \eqref{guideq} do nothing to stop you from picking whatever global phase you'd like at each time.)  In general, some of these wave functions will satisfy the Schr\"{odinger} equation \eqref{schrod} and others will not.  To constrain the time-dependence of the phase when using a wave function to describe histories, \eqref{thetaT} can be imposed as a third link between the wave function and the particles (in addition to \eqref{probcon} and \eqref{guideq}).  Because \eqref{probcon}, \eqref{guideq}, \eqref{hamster}, and \eqref{thetaT} hold, the wave function must obey the Schr\"{o}dinger equation.

This section began with the observation that there are states in Newtonian QM that cannot be described by a wave function.  However, these can be excised by imposing the Quantization Condition.  Given a state that \emph{can} be described by a wave function, one might hope that this wave function would be unique.  Sometimes it is not.  A wave function aptly describes a state in Newtonian QM \emph{at a time} if \eqref{probcon} and \eqref{guideq} are satisfied.  But, if these are the only constraints, a \emph{history} in Newtonian QM can always be described by many wave functions.  So, there is freedom to add a third connection between the wave function and the particles.  Imposing \eqref{thetaT} proves a convenient choice as it guarantees that the wave function obeys the Schr\"{o}dinger equation---a desirable feature since the point of introducing a wave function was to clarify the connection between Newtonian QM and standard treatments of quantum mechanics.

Because a wave function can be introduced to describe the world density and the velocity fields, one is free to use well-known techniques to calculate the time evolution of the wave function and use that to determine how the world density and velocity fields evolve.  However, there is evidence that it is sometimes easier to use the trajectories of worlds to calculate the time evolution \citep{wyatt2005, HDW}.

\section{Probability: Versus Everettian Quantum Mechanics}\label{probEQM}

The Born Rule is easier to justify in Newtonian QM than in the many-worlds interpretation.  In Everettian QM, there is dispute over how one can even make sense of assigning probabilities to measurement outcomes when the way the universe will branch is deterministic and known (the incoherence problem).  There is also the quantitative problem of why the Born Rule probabilities are the right ones to assign.  Recent derivations tend to appeal to complex decision-theoretic arguments, which, although they may ultimately be successful, are not uncontroversially accepted \citep{MWbook}.  Things look worrisome because there are some \emph{prima facie} plausible ways of counting agents which yield the result that \emph{the vast majority of agents} see relative frequencies of experimental outcomes which deviate significantly from those predicted by the Born Rule (although the total amplitude-squared weight of the branches in which agents see anomalous statistics is small).  Newtonian QM does not run into similar problems since the number of worlds in a particular region of configuration space is always proportional to $|\Psi|^2$.  At any time, most agents are in high amplitude regions.  So, in typical measurement scenarios, most agents will see long-run frequencies which agree with the predictions of the Born Rule.

Were a proponent of Prodigal QM to claim similar advantages over Everettian QM, one could reasonably object that the Born Rule is recovered only because it was put in by hand.  In Prodigal QM, \eqref{probcon} is an additional postulate.  In Newtonian QM, it is not.  The density of worlds is given by $|\Psi|^2$ because $\Psi$ is definitionally related to the density of worlds by \eqref{probcon}  (see \textsection \ref{defining}).  The wave function is, after all, not fundamental but a mere description of $\rho$ and the $\vv{v}_k$s.

\section{Probability: Versus Bohmian Mechanics}\label{probBOHM}

Although it is widely agreed that the Born Rule can be justified in Bohmian mechanics, there is disagreement about how exactly the story should go.  In this section I will briefly discuss three ways of justifying the Born Rule in Bohmian mechanics and then argue that Newtonian QM can give a cleaner story.  First, though, note an important similarity between the two theories.  According to Newtonian QM each world follows an approximately Bohmian path through configuration space.  So if you think that worlds in which particles follow Bohmian trajectories are able to reproduce the results of familiar quantum experiments, you should think worlds in Newtonian QM can too.

In Bohmian mechanics, not all initial conditions reproduce the statistical predictions of quantum mechanics.  That is, not all specifications of the initial wave function $\Psi(0)$ and particle configuration $(\vv{x}_1(0), \vv{x}_2(0), ...)$ yield a universe in which experimenters would see long-run statistics of measurements on subsystems which agree with the predictions of the Born Rule.  Why should we expect to be in one of the universes with Born Rule statistics?  One way to respond to this problem is to add a \emph{postulate} to the theory which ensures that ensembles of particles in the universe will (or almost certainly will) display Born Rule statistics upon measurement (e.g., \citealp[][\textsection 3.6.3]{holland}).  A second option is to argue that \emph{typical} universes are such that Born Rule frequencies will be observed when measurements are made \citep{durr1992}.  To say that such results are ``typically'' observed is to say that: for any initial wave function $\Psi(0)$, the vast majority of initial particle configurations reproduce Born Rule statistics.  Speaking of the ``vast majority'' of initial configurations only makes sense relative to a way of measuring the size of regions of configuration space; here the measure used is given by $|\Psi|^2$.  A third option: one could argue that many initial states will start to display Born Rule statistics sufficiently rapidly that, since we are not at the beginning of the universe, we should expect to see Born Rule frequencies now even if such frequencies were not displayed in the distant past \citep{valentini2005}.

Each of these proposals faces challenges.  The additional postulates which might be added to the theory look ad hoc.  The measure used to determine typicality must be satisfactorily justified.\footnote{For a statement of the objection, see \citet[\textsection 5.4]{dickson1998}.  For a variety of reasons to regard the measure as natural, see \citet{struyvegoldstein}.}  The desirable evolution of states described in the third option has only been demonstrated in relatively simple cases.  Also, there will certainly exist initial conditions that do not come to display Born Rule statistics sufficiently rapidly and these must somehow be excluded.  To the extent that one finds these objections to Bohmian strategies worrisome, it is an advantage of the new theory that it avoids them.

Although Newtonian QM, like Bohmian mechanics, permits a particular world to have a history of measurement results where the frequencies of outcomes do not match what one would expect from the Born Rule, it is \emph{impossible} for the density of worlds to deviate from $|\Psi|^2$.  So, in light of the results in \citet{durr1992}, it will always be the case that Born Rule statistics are observed in the vast majority of \emph{worlds} in any \emph{universe} of Newtonian QM.  Since we're not sure which world we are in, we should expect to be in one in which Born Rule statistics are observed.

\section{Probability: Newtonian QM}\label{probNQM}

If the universe's evolution is deterministic and the initial state is known, what is there left for an agent to assign probabilities to?  There is no incoherence problem in Newtonian QM since, given the state of the universe, one is generally uncertain which of the many distinct worlds one is in.  There will always be many possibilities consistent with one's immediate experiences.  The uncertainty present here is \emph{self-locating uncertainty} \citep[see][]{lewis1979}.  Of course, there will generally also be uncertainty about the state of the universe.

On to the quantitative problem:\footnote{See also the discussion in \citet[\textsection 2.4]{bostrom2012}.} Given a particular distribution of worlds $\rho$ and set of velocity fields $\vv{v}_k$, that is, given a specification of the state of the universe, one ought to assign \emph{equal credence} to being in any of the worlds consistent with one's evidence.\footnote{This follows from a more general epistemic principle defended in \citet{elga2004}.}  Because there are only a finite number worlds, this advice is unambiguous.  As it turns out, this basic indifference principle suffices to derive the correct quantum probabilities.  Consider an idealized case in which the agent knows the world density and the velocity fields, and knows that there is an agent in each of these worlds having experiences indistinguishable from her own.  In this case, the above indifference principle tells her to assign probabilities to being in different regions of configuration space in accordance with $\rho$.  Since $\rho=|\Psi|^2$, she must assign credences in accordance with $|\Psi|^2$ and thus in agreement with the Born Rule.  Next, suppose this agent learns the outcome of an experiment.  Then she ought to assign zero credence to the worlds inconsistent with her evidence and reapportion that credence among those which remain (keeping the probability of each non-eliminated world equal).  This updating is analogous to learning which branch you are on after a measurement in Everettian QM.

In general, the probability agent $S$ ought to assign to her own world having property $A$, conditional on a particular state of the universe at a certain time, is
\begin{align}
\mbox{Pr}\big(A\big|\rho,\vv{v}_1, \vv{v}_2, ...)&=\frac{\mbox{\# of worlds with property }A\mbox{ and a copy of }S}{\mbox{\# of worlds with a copy of }S}
\nonumber
\\
&=\frac{\bigintsss\!\! d V_{AS} \ \rho(\vv{x}_1,\vv{x}_2,...)}{\bigintsss\!\! d V_S \ \rho(\vv{x}_1,\vv{x}_2,...)}=\frac{\bigintsss\!\! d V_{AS} \ |\Psi(\vv{x}_1,\vv{x}_2,...)|^2}{\bigintsss\!\! d V_S \ |\Psi(\vv{x}_1,\vv{x}_2,...)|^2}\ .
\end{align}
Here $A$ could be something like, ``the pointer indicates 7'' or ``the particle just fired will hit in the third band of the interference pattern.''  The volume $V_{S}$ delimits the set of worlds, specified by a region of configuration space, compatible with $S$'s data.  Worlds in this region are such that previous experiments had the outcomes $S$ remembers them having, macroscopic arrangements of particles match what $S$ currently observes, and some person is having the same conscious experiences as $S$.\footnote{For simplicity, I have neglected the possibility that $S$'s memories or current observations are deceptive.}  The volume $V_{AS}$ gives the set of worlds compatible with $S$'s data in which $A$ holds.\footnote{Note that the boundaries of $V_{S}$ and $V_{AS}$ will often depend on $\rho$ and the $\vv{v}_k$s.}  These conditional probabilities can be used to test hypotheses about $\rho$ and the $\vv{v}_k$s and thus to learn about the state of the universe (not just one's own world) from experience.

\section{Continuous Infinity or Mere Multitude of Worlds?}\label{multitude}

So far, we have taken $\rho$ to describe the distribution of a large but finite number of worlds.  But, one might be tempted to defend a variant of Newtonian QM in which there are a continuous infinity of worlds, one at every point at which $\rho$ is non-zero.  This causes trouble.  The \emph{meaning} of $\rho$ becomes unclear if we move to a continuous infinity of worlds since we can no longer understand $\rho$ as yielding the \emph{proportion} of all worlds in a given volume of configuration space upon integration over that volume.  There would be infinite worlds in any finite volume (where $\rho\neq 0$) and infinite total worlds.  If $\rho$ doesn't give the proportion of worlds in a region, it is unclear why epistemic agents should apportion credences as recommended in the previous section.  So, the continuous variant, if sense can be made of it, faces the quantitative probability problem head on.

As discussed in \textsection \ref{badstates}, the dynamical law proposed for Newtonian QM \eqref{eom} is not fundamental.  If it somehow turns out that we cannot view the force caused by the quantum potential as arising from an interaction between individual worlds, this would provide a reason to accept a continuous infinity of quantum worlds over a mere multitude.  It might appear to be a strength of the continuous variant that its laws can already be precisely stated, but I expect that this advantage will evaporate when possible fundamental interactions are formulated for the discrete variant.  The continuous variant does have a serious advantage:  the continuum approximation (\textsection \ref{badstates}) is no approximation.  Particles will unerringly follow Bohmian trajectories.

\section{Ontology}\label{ontology}

According to Newtonian QM, what the universe contains is a finite collection of worlds.  There are at least two ways to precisify this idea.  First, one might take \emph{configuration space} to be the fundamental space, inhabited by point-particles (worlds).  Second, one might take the fundamental space to be ordinary \emph{three-dimensional space}, inhabited by particles in different worlds.

According to the first picture, on the fundamental level, the universe is 3$N$-dimensional and contains a large number of point particles, each of which has dynamics so complex that it merits the name of ``world'' or ``world-particle.''  Forces between these world-particles are Newtonian and the dynamics are local.  Here Newtonian QM is a theory of the Newtonian dynamics of a fluid of world-particles in 3$N$-dimensional space.  \citet{albert1996} has argued that the one world of Bohmian mechanics can be understood as a world-particle which moves around in configuration space guided by the wave function.  He provides a way of explaining how the appearance of a three-dimensional world arises from the motion of this world-particle which applies \emph{mutatis mutandis} to Newtonian QM in which there are more world-particles executing the same old Bohmian dances.

On the second picture there are particles interacting in three-dimensional space, nothing more.\footnote{This second option resembles the novel ontology for the many-worlds interpretation proposed by \citet{allori2011}.}  Space is very densely packed with particles, but not all particles are members of the same world.  Some particles are members of world \#1, some of world \#2, etc.  What world a particle belongs to might be a primitive property, like its mass or charge.  The equation of motion for a particle in world \#827, \eqref{eom}, says that the force from the potential $V$ depends only on the positions of the other particles in world \#827.  However, the quantum potential introduces an inter-world force whereby particles that are not members of world \#827 can still impact the trajectory of a particle in this world.  So, particles which happen to be members of the same world interact in one way, whereas particles which are members of different worlds interact another way. 
 
In the many-worlds interpretation, one must tell a somewhat complicated story about how people and quantum worlds arise as emergent entities in the time-evolving quantum state \citep[e.g.,][]{wallace2003}.  This story may not be successful.  It might be the case that wave functions evolving in accordance with the Schr\"{o}dinger equation are incapable of supporting life or at least lives that feel like ours \citep{maudlin2010}.  If that's right, Newtonian QM has a potential advantage.  On the second ontological picture, people are built from particles in the usual way.  On the first ontological picture, there is a story about emergence that must be told but the details of the story are very different from the Everettian one and it succeeds or fails independently.

If, on the other hand, the Everettian story about emergence is successful, then Bohmian mechanics (as formulated here) faces the \emph{Everett-in-denial objection} \citep{deutsch1996lockwood,brown2005,valentini2010}.  Both Everettian QM and Bohmian mechanics contain in their fundamental ontology a wave function which always obeys the Schr\"{o}dinger equation.  If such a wave function is sufficient for there to be creatures experiencing what appears upon not-too-close inspection to be a classical world, then Bohmian mechanics, like Everettian QM, includes agents who see every possible outcome of a quantum measurement.  If the Everettian story about emergence works and the Everett-in-denial objection against Bohmian mechanics is successful, then Newtonian QM has a serious advantage over Bohmian mechanics.  Newtonian QM cannot be accused of being a many worlds theory in disguise since the theory embraces its many worlds ontology.

\section{Symmetries: Time Reversal and Galilean Boosts}\label{sym}

Newtonian QM can help us understand symmetry transformations in quantum mechanics.  First, consider time reversal.  \citet{albert2000} proposes an intuitive and general account of time reversal symmetry in physical theories which judges QM, in all of its familiar precisifications, to \emph{fail} to be time-reversal invariant.  A deterministic physical theory specifies which sequences of instantaneous states are allowed and which are forbidden through dynamical laws.  If the laws allow the time-reversed history of instantaneous states for any allowed history of instantaneous states, then the theory is deemed time-reversal invariant.  In theories like Bohmian mechanics or Everettian QM, the instantaneous state includes the wave function at a time $\Psi(\vv{x}_1,\vv{x}_2,...,t)$ and a complete history includes the wave function at all times.  The time reverse of the history is $\Psi(\vv{x}_1,\vv{x}_2,...,-t)$.  $\Psi(\vv{x}_1,\vv{x}_2,...,-t)$ will not necessarily satisfy the Schr\"{o}dinger equation whenever $\Psi(\vv{x}_1,\vv{x}_2,...,t)$ does---so quantum mechanics is judged not to be time-reversal invariant.  However, $\Psi^*(\vv{x}_1,\vv{x}_2,...,-t)$ will \emph{always} satisfy the Schr\"{o}dinger equation whenever $\Psi(\vv{x}_1,\vv{x}_2,...,t)$ does (standard textbook accounts take this to be the time reversed history and thus judge the theory to be time-reversal invariant).

In Newtonian QM, it is straightforward to show that time reversing the history of particle trajectories amounts to changing the history of the wave function from $\Psi(\vv{x}_1,\vv{x}_2,...,t)$ to $\Psi^*(\vv{x}_1,\vv{x}_2,...,-t)$.  The \emph{instantaneous} state of the world is specified by giving the locations (but \emph{not} the velocities) of all of the particles in all of the worlds.  The time reversal operation thus takes the history $\rho(\vv{x}_1,\vv{x}_2,...,t)$ and $\vv{v}_k(\vv{x}_1,\vv{x}_2,...,t)$ to $\rho(\vv{x}_1,\vv{x}_2,...,-t)$ and $-\vv{v}_k(\vv{x}_1,\vv{x}_2,...,-t)$.  By \eqref{guideq2}, flipping the phase generates a wave function which describes the flipped velocities of particles in the time-reversed history.  The complex conjugation in the textbook time reversal operation for quantum mechanics can be explained as deriving from a reversal in the velocities of the particles.

Newtonian QM is time-reversal invariant according to Albert's account.  Even if one doesn't agree with Albert's account of time-reversal invariance, it is a virtue of this theory over others that it can give a simple explanation of \emph{why} the wave function transforms in the textbook way under time-reversal.

Next, consider Galilean boosts.  In a similar spirit to Albert's criticism of the standard account of time-reversal, one could argue that quanrtum mechanics is not invariant under Galilean boosts since the equations of motion are not generally obeyed when we take $\Psi(\vv{x}_1,\vv{x}_2,...,t)$ to $\Psi(\vv{x}_1-\vv{w}t,\vv{x}_2-\vv{w}t,...,t)$.\footnote{A point made by Albert in presentations.  See also \citet{valentini1997}.}  The invariance of quantum mechanics under Galilean boosts is sometimes demonstrated by showing that, for certain potentials, there \emph{exists} a transformation of the state which appropriately shifts the probability density and guarantees satisfaction of the Schr\"{o}dinger equation (e.g., \citealp[\textsection 4.3]{ballentine}).  Under a boost by $\vv{w}$, the wave function is supposed to transform as
\begin{equation}
\Psi_0(\vec{x}_1,\vv{x}_2,...,t)\stackrel{\vv{w}}{\longrightarrow}\Psi(\vec{x}_1,\vv{x}_2,...,t)= e^{\frac{i}{\hbar}\sum_k{\left\{m_k \vv{w}\cdot\vv{x}_k-\frac{1}{2}m_k |\vv{w}|^2 t\right\}}}\Psi_0(\vv{x}_1-\vv{w}t,\vv{x}_2-\vv{w}t,...,t)\ .
\label{transf}
\end{equation}
It's interesting that there exists a transformation which moves probability densities in the right way and guarantees that the Schr\"{o}dinger equation is invariant under boosts, but it is unclear why this particular transformation is the one that really represents Galilean boosts.  In Newtonian QM this transformation of the wave function results from boosting the velocities of all of the particles in all of the worlds.

Adding $\vv{w}$ to the velocity of each particle transforms the original density $\rho_0(t)$ and the original velocity fields $\vv{v}_{0k}(t)$ to
\begin{align}
\rho(\vv{x}_1,\vv{x}_2,...,t) &= \rho_0(\vv{x}_1-\vv{w}t,\vv{x}_2-\vv{w}t,...,t)
\nonumber
\\
\vv{v}_{k}(\vv{x}_1,\vv{x}_2,...,t) &= \vv{v}_{0k}(\vv{x}_1-\vv{w}t,\vv{x}_2-\vv{w}t,...,t)+\vv{w}\ .
\label{newstuff}
\end{align}
Suppose $\Psi_0(t)$, $\rho_0(t)$, and the $\vv{v}_{0k}(t)$s satisfy \eqref{probcon}, \eqref{guideq}, and \eqref{thetaT}; that is, $\Psi_0(t)$ \emph{describes} this density and these velocity fields.  Then, the new wave function $\Psi(t)$ generated by the transformation in \eqref{transf} will satisfy \eqref{probcon}, \eqref{guideq}, and \eqref{thetaT} for the $\rho(t)$ and $\vv{v}_k(t)$s in \eqref{newstuff}, provided that the potential $V$ is translation invariant (as the reader can verify).  Thus, \eqref{transf} gives a general recipe for finding a wave function which correctly describes the boosted particles.

\section{Spin-1/2 Particles}\label{withspin}

There appears to be serious trouble on the horizon for this new theory.  In Bohmian mechanics spin is often treated as a property of the wave function, not the particles pushed along by it.\footnote{e.g., \citet[\textsection 8.4]{durr2009} and \citet[ch. 7]{albert1994}}  So, if we remove the wave function, it looks like we'll lose all of the information about the spin of the system!  Actually, there is a very natural way to extend Newtonian QM to a single particle with spin.  If we endow the particle with a \emph{definite} spin in every world, we can recover the standard dynamics.  Here I'll apply to Newtonian QM a strategy which has been used in quantum hydrodynamics and (a version of) Bohmian mechanics (see \citealp[][ch. 9]{holland} and references therein).

Consider the dynamics of a single spin-$1/2$ particle.  To our basic ontology, consisting of a distribution of worlds $\rho(\vv{x},t)$ where the particle has velocity $\vv{v}(\vv{x},t)$ in each world, let us add a property to the particle in each world: spin magnetic moment.  The spin magnetic moment $\vv{u}(\vv{x},t)$ of a particle can be specified by a polar angle $\alpha(\vv{x},t)$, an azimuthal angle $\beta(\vv{x},t)$, and a constant $\mu$ (for an electron, $\mu\approx\frac{-e \hbar}{2 m}$, where $e$ is the magnitude of the electron's charge).
\begin{equation}
\vv{\mu}=\mu \left(\begin{array}{c}\sin\alpha\cos\beta \\ \sin\alpha\sin\beta \\ \cos \alpha \end{array} \right)
\label{mumu}
\end{equation}
Alternatively, we can speak of the particle's internal angular momentum $\vv{S}$, which is related to $\vv{\mu}$ by
\begin{equation}
\vv{S}=\frac{\hbar}{2 \mu} \vv{\mu}\ .
\label{crispy}
\end{equation}
With the magnetic moment in hand, we can partially define\footnote{This definition is only partial as $\theta$ is left unspecified.} the spinor wave function $\chi$ from $\rho$ and $\vv{\mu}$ by
\begin{equation}
\chi=\left(\begin{array}{c}\chi_+ \\ \chi_- \end{array}\right)=\left(\begin{array}{c} \sqrt{\rho}\cos\frac{\alpha}{2} \: e^{i\theta} \\ \sqrt{\rho}\sin\frac{\alpha}{2} \: e^{i\theta+i\beta} \end{array}\right)\ ,
\label{spinor}
\end{equation}
similar to \eqref{first}.  Here the $z$-spin basis is used to represent the spinor.

The Bohmian guidance equation for a spin-$1/2$ particle is
\begin{equation}
\vv{v}=\frac{\hbar}{m}\mbox{Im}\left[\frac{\chi^\dagger \vv{\nabla} \chi}{\chi^\dagger \chi}\right]\ .
\label{}
\end{equation}
Inserting the expression for $\chi$ in \eqref{spinor} yields
\begin{equation}
\vv{v}=\frac{\hbar}{m}\vv{\nabla}\theta+\frac{\hbar}{m}\sin^2\frac{\alpha}{2}\: \vv{\nabla}\beta\ ,
\label{guideq3}
\end{equation}
similar to \eqref{guideq2}.

The Pauli equation for a spin-$1/2$ particle in the presence of an external magnetic field is
\begin{equation}
i\hbar\frac{\partial}{\partial t} \chi = \left\{\frac{-\hbar^2}{2 m}\nabla^2+V-\mu \vv{B} \cdot \sigma\right\}\chi\ ,
\label{RingerSchrodinger}
\end{equation}
where $\sigma$ are the Pauli spin matrices.  To focus on spin, the contributions to the Hamiltonian arising because the particle has a charge (not just a magnetic moment) have been omitted.  From \eqref{spinor}, \eqref{guideq3}, and \eqref{RingerSchrodinger} one can derive the time dependence of $\vv{\mu}$ and $\vv{v}$.  The magnetic moment vector evolves as
\begin{align}
\frac{\hbar}{2 \mu}\frac{d\vv{\mu}}{dt}=&\frac{\hbar^2}{4 m \mu^2 \rho} \vv{\mu} \times \left[ \partial_a \left(\rho \: \partial_a \vv{\mu}\right) \right] + \vv{\mu} \times \vv{B}
\nonumber
\\
\frac{d \vv{S}}{dt}=&\vv{\mu} \times \vv{B}_{\text{Tot}}\ ,
\label{muT}
\end{align}
using the Einstein summation convention over spatial index $a$.\footnote{Result as in \citet[eq.\ 9.3.15]{holland}.  Note that different conventions are adopted for the sign of $\mu$.  \eqref{eom2} is in agreement with Holland's eq.\ 9.3.19, although written in a more suggestive form.}  The right hand side gives the net torque on the particle, which arises from a quantum and a classical contribution.  These torques can be combined by defining
\begin{equation}
\vv{B}_{\text{Tot}}\equiv\vv{B}+\frac{\hbar^2 \left[ \partial_a \left(\rho \: \partial_a \vv{\mu}\right) \right]}{4 m \mu^2 \rho}\ .
\label{bagnetic}
\end{equation}
The net magnetic field $\vv{B}_{\text{Tot}}$ is the sum of a classical and a quantum contribution.  \eqref{muT} gives the classical dynamics for the angular momentum of a magnetic dipole in the presence of the magnetic field $\vv{B}_{\text{Tot}}$.

From \eqref{spinor}, \eqref{guideq3}, and \eqref{RingerSchrodinger}, it follows that the acceleration can be expressed as
\begin{align}
m \vv{a}=-\vv{\nabla}\left[Q+Q_{P}+V\right]+\mu_a \vv{\nabla} {B_{\text{Tot}}}_a\ .
\label{eom2}
\end{align}
This is simply the equation of motion for a particle without spin \eqref{eom} with two new terms: the classical force on a particle with magnetic moment $\vv{\mu}$ from a magnetic field $\vv{B}_{\text{Tot}}$ and a spin-dependent contribution to the quantum potential,
\begin{equation}
Q_P=\frac{\hbar^2}{8 m \mu^2}\vv{\mu}\cdot\left(\nabla^2 \vv{\mu}\right)=\frac{1}{2 m}\vv{S}\cdot\left(\nabla^2 \vv{S}\right)\ .
\end{equation}
As with the quantum potential $Q$ discussed in \textsection \ref{removing}, this new term represents an interaction between worlds (as does the quantum contribution to the net magnetic field $\vv{B}_{\text{Tot}}$).  Together, the above equations of motion for $\vv{\mu}$ and $\vv{v}$,  \eqref{muT} and \eqref{eom2}, serve to define Newtonian QM for a single spin-$1/2$ particle.  We can omit any mention of the spinor wave function $\chi$ or the phase $\theta$ in the fundamental laws.  The equations of motion for $\vv{\mu}$ and $\vv{v}$, which govern the evolution of $\rho$ via \eqref{hamster}, will guarantee that $\rho$, $\vv{\mu}$, and $\vv{v}$ will evolve as if they were governed by a spinor wave function satisfying the Pauli equation, provided that the velocity field obeys a constraint like the one imposed for spin-$0$ particles in \textsection \ref{defining},
\begin{equation}
\oint{\left(m\vv{v}-\hbar \sin^2\frac{\alpha}{2} \: \vv{\nabla}\beta \right)\cdot d\vv{\ell}}=nh\ .
\label{}
\end{equation}

In Newtonian QM, particles have well-defined spin magnetic moments at all times.  How can the theory recover the results of standard experiments involving spin if particles are never in superpositions of different spin states?  Consider, for example, a $z$-spin ``measuring'' Stern-Gerlach apparatus.  Suppose the wave function is in a superposition $z$-spin up and $z$-spin down: $\frac{1}{\sqrt{2}}\left|\uparrow_z\right\rangle+\frac{1}{\sqrt{2}}\left|\downarrow_z\right\rangle$.  When passed through the inhomogeneous magnetic field, the wave function will split in two.  On the standard account, the particle will be found in either the upper region (corresponding to $z$-spin up) or the lower region (corresponding to $z$-spin down) upon measurement with equal probability.  In Newtonian QM, there is initially an ensemble of worlds, in each of which the particle has some initial position in the wave packet and in all of which the particle's spin magnetic moment points squarely in the $x$-direction.  A particle in the top half of the initial wave packet has its spin rotated to point in the $z$-direction as it passes through the Stern-Gerlach apparatus (in accordance with \eqref{muT}); a particle in the lower portion will end up with spin pointing in the negative $z$-direction.  In this theory, the Stern-Gerlach apparatus does not \emph{measure} $z$-spin, but instead forces particles to align their magnetic moments along the $z$-axis.  This is also how Stern-Gerlach measurements are interpreted in versions of Bohmian mechanics where particles have definite spins \citep[see][ch. 9]{dewdney1986,holland}.

\section{Conclusion}\label{conchshell}

\emph{An optimistic synopsis}: Once we realize that Newtonian QM is a viable way of understanding non-relativistic quantum mechanics, we see that we never needed to overthrow Newtonian mechanics with a quantum revolution.  One can formulate quantum mechanics in terms of point particles interacting via Newtonian forces.  The mysterious wave function is merely a way of summarizing the properties of particles, not a piece of fundamental reality.

There are a variety of reasons not to like this theory.  First, there is arguably a cost associated with the abundance of other worlds which, although detectable via their interactions with our own world, are admittedly odd.  Second, the space of states for the theory is larger than one might like in two distinct ways:  There are possible combinations of $\rho$ and the $\vv{v}_k$s that do not correspond to any wave function because the velocity fields cannot be expressed as the gradient of a phase (\textsection \ref{defining}).  There are also states of the universe where the number of worlds is not sufficiently large for the continuum description to be valid (\textsection \ref{badstates}).  Even if there are a great many worlds, slight divergence from the predictions of standard quantum mechanics is to be expected.  Third, it is a shortcoming of the current formulation of Newtonian QM that we must approximate the actual distribution of worlds as continuous and cannot yet formulate the fundamental equation of motion precisely for a discrete collection of worlds (\textsection \ref{badstates}).  Finally, the theory is limited in that it is not here extended to systems of multiple particles with spin or to relativistic quantum physics.

In addition to its seductive conservatism, I view the following comparative strengths as most compelling.  Against the many-worlds interpretation, Newtonian QM has two main advantages.  First, there is no incoherence problem or quantitative probability problem---the Born Rule can be justified quickly from self-locating uncertainty (\textsection \ref{probEQM}).  Second, the theory avoids the need to explain how worlds emerge from the wave function---worlds are taken to be fundamental (\textsection \ref{ontology}).  Compared to Bohmian mechanics, the theory is arguably simpler---it replaces an ontology of wave functions and particles with one just containing particles (\textsection \ref{removing}).  Newtonian QM's explanation of why we should expect our world to reproduce Born Rule statistics is potentially more compelling than the Bohmian stories (\textsection \ref{probBOHM}).  Also, Newtonian QM is forthright about its many worlds character, sidestepping the Everett-in-denial objection (\textsection \ref{ontology}).\\

\textbf{Acknowledgements}
Thanks to David Baker, Gordon Belot, Cian Dorr, Detlef D\"{u}rr, J. Dmitri Gallow, Sheldon Goldstein, Michael Hall, Daniel Peterson, Laura Ruetsche, Ward Struyve, Nicola Vona, and two anonymous referees for very useful feedback on drafts of this article.  Thank you to Adam Becker, Sean Carroll, Dirk-Andr\'{e} Deckert, Neil Dewar, Benjamin Feintzeig, Sophie Monahan, Cat Saint-Croix, Jonathan Shaheen, and Howard Wiseman for helpful discussions.  This material is based upon work supported by the National Science Foundation Graduate Research Fellowship under Grant No. DGE 0718128.

\bibliography{QMasCPbibfile2}

\end{document}